 \documentclass[final,3p,times]{elsarticle}

\usepackage{amssymb,amsmath,bm}

\begin{document}

\renewcommand{\labelenumi}{(\roman{enumi})}

\begin{frontmatter}

\title{A note on the general multi-moment constrained flux reconstruction formulation for high order schemes}

\author[titech]{Feng Xiao }\corref{cor}
\author[uosaka]{Satoshi Ii}
\author[xju]{Chungang Chen}
\author[cma]{Xingliang Li}

\address[titech]{Department of energy sciences, Tokyo Institute of Technology,   
4259 Nagatsuta, Midori-ku, Yokohama, 226-8502, Japan}
\address[uosaka]{ Department of Mechanical Science \& Bioengineering, Osaka University, Toyonaka, Osaka, 560-8531, Japan}
\address[xju]{School of Human Settlement and Civil Engineering, Xi'an Jiaotong University, 
28 Xianning West Road, Xi'an, 710049, China }
\address[cma]{Center of Numerical Weather Prediction, China 
                Meteorological Administration, 46 Zhongguancun South street, Beijing, 100081, China}
\cortext[cor]{Corresponding author: Feng Xiao, \it{xiao@es.titech.ac.jp}}

\begin{abstract}
This paper presents a general formulation to construct high order numerical schemes by using multi-moment constraint conditions on the flux function reconstruction. The new formulation, so called multi-moment constrained flux reconstruction (MMC-FR), distinguishes itself essentially from the flux reconstruction formulation (FR) of Huynh (2007) by imposing  not only the continuity constraint conditions on the flux function at the cell boundary, but also other types constraints which may include those on the spatial derivatives or the point values. This formulation can be also interpreted as a blend of Lagrange interpolation Hermite interpolation, which provides a numerical framework to accommodate a wider spectrum of high order schemes.  Some representative schemes will be presented and evaluated through Fourier analysis and numerical tests.
\end{abstract}

\begin{keyword}
High order scheme \sep flux reconstruction \sep multi-moment constraint \sep nodal formulation \sep derivative Riemann problem \sep conservation

\end{keyword}

\end{frontmatter}


\section{Introduction}
 
Solving governing equations point-wisely at the specified nodes (solution points) results in a class of very efficient schemes, which are usually known as nodal form, differential form or collocation form, in which no explicit numerical quadrature is involved. Examples of this kind are the  nodal discontinuous Galerkin (DG) method\cite{hw08} and the spectral collocation method\cite{ko96}. In \cite{hu07}, Huynh suggested a general formulation, so called flux reconstruction (FR), from which the aforementioned schemes among others can be retrieved from a Lagrange interpolation polynomial with different correction functions that ensure the continuity of the cell-wisely constructed flux functions at cell interfaces. The DG method can be derived if Radau or Legendre polynomial is chosen for the correction function, while the spectral collocation method can be retrieved if the correction function is collocated at the Chebyshev points to a zero function inside the cell while matches the modified flux function to the continuous value at the cell boundary. These schemes minimize the modification to the primary Lagrange interpolation, and only requires  the reconstructed flux function itself to be continuous at the cell boundary, which is the necessary condition for the conservation. Wang and Gao have implemented the FR in unstructured meshes \cite{wg09}, and Vincent et al. have devised some stable subset schemes of Huynh's FR formulation with a parameter switching \cite{vcj11-1,vcj11-2}. 

If we interpret the flux reconstruction as an interpolation procedure with some constraining conditions given, the reconstruction of the modified flux function in Huynh's FR formulation can be essentially viewed as a Lagrange interpolation procedure, i.e. all the constraints that lead to the final modified flux function are solely the point values (PV) including those at the cell boundaries for conservation and stability and those at the solution points.     

In an alternative direction, we have so far explored the possibility to make use of the Hermite interpolation in constructing high order schemes\cite{ya01b,xi01,xi04,ii07,ii09}. Compared to the Lagrange interpolation where only PV is used, the Hermite interpolation uses also the spatial derivatives and even integral as the constraints to determine the reconstruction interpolation. Hence, we categorize this class as ``multi-moment" type schemes against the formulations where only one kind ``single moment", for example the PV in the Lagrange interpolation, is used. In a recent paper \cite{ii09}, we proposed a nodal type formulation, so called multi-moment constrained finite volume method (MCV), where the PVs at solution points are updated by evolution equations which are derived via a Hermite interpolation reconstruction under the multi-moment constraints including point values, derivatives and integral. The MCV method has been implemented for unstructured grids in spherical geometry with competitive performance in numerical accuracy and computational efficiency \cite{ii10,chen12}. Using multi-moment constraints provides another platform to construct high-order schemes. 

We in this paper presents a general formulation that makes use of not only the PV but also other moments like derivatives as the constraints to build the modified flux function. The formulation is called multi-moment constrained flux reconstruction (MMC-FR), from which new high order schemes can be straightforwardly devised, such as the MCV method. 

This paper is organized as follows. The multi-moment constrained flux reconstruction formulation is presented in section 2 with comparisons to Huynh's FR formulation. A von Neumann analysis and the convergence rate evaluation using grid refinement tests are given in section 3 for some representative schemes of MMC-FR approach, and the paper ends with some remarks in section 4.

\section{Multi-moment constrained flux reconstruction formulation}
We consider the following conservation law  
\begin{equation}
{{\partial u} \over {\partial t}}+{{\partial f} \over {\partial x}} =0,  
\label{1dge}
\end{equation}
where  $u$  is the solution function, and  $f(u)$ the flux function.


The computational domain is divided into $I$ non-overlapping cells or elements $\Omega_i=[x_{i-{1\over2}},x_{i+{1\over2}}]$, $i=1,2,\cdots,I$, and $K$ solution points $x_{ik}$, $k=1,2,\cdots,K$, are set over  $\Omega_i$ where the solution $u_{ik}$, $k=1,2,\cdots,K$, is computed.  Suppose that a proper approximation of the flux function  $\tilde{f}_i(x)$ is constructed, we can immediately update the solutions within  $\Omega_i$ by the following point-wise semi-discretized equations at solution points $x_{ik}$, 
\begin{equation}
{{d u_{ik}} \over {d t}}=-\left [ {{d \tilde{f}_i(x)} \over {d x}}\right ]_{ik}, \quad k=1,2,\cdots,K.  
\label{upu}
\end{equation}
The above equation \eqref{upu} is of nodal form, also termed as differential form or strong form, which covers a wide range of schemes including the collocation methods. The central task left now is how to reconstruct the flux function $\tilde{f}_i(x)$. In principle, the way to reconstruct  $\tilde{f}_i(x)$ makes  difference among the numerical schemes.  

Given the solution $u_{ik}$ at $x_{ik}$, $k=1,2,\cdots,K$, a piece-wise Lagrange interpolation polynomial of degree $K-1$ for $\Omega_i$ reads
\begin{equation}
u_i(x)=\sum^K_{k=1}u_{ik}\phi_{ik}(x),
\label{lagintp}
\end{equation}
where $\phi_{ik}(x)$ is the Lagrange basis function,   
\begin{equation}
\phi_{ik}(x)=\prod^K_{l=1,l\neq k}\frac{x-x_{il}}{x_{ik}-x_{il}}.
\label{lagintp_u}
\end{equation}

Given flux $f(u)$ as the function of solution $u$, we have the flux function $f_{ik}$ at the solution points $x_{ik}$, $k=1,2,\cdots,K$, simply by $f_{ik}=f(u_{ik})$. The piece-wisely reconstructed polynomial for flux function is then obtained as 
\begin{equation}
f_i(x)=\sum^K_{k=1}f_{ik}\phi_{ik}(x),  
\label{lagintp_f}
\end{equation}
which is of degree $K-1$, same as that for the solution function $u_i(x)$.  

We call \eqref{lagintp_f} the primary reconstruction which is separately constructed over each cell and thus broken from cell by cell. The primary reconstruction cannot be directly used to calculate \eqref{upu},  further modification is required to ensure at least the $C^0$ continuity at the two ends ($x_{i-{1\over2}}$ and $x_{i+{1\over2}}$) of cell $\Omega_i$, which is the necessary condition for numerical conservation and computational stability. 

The way to reconstruct the modified flux function $\tilde{f}_i(x)$ in an MMC-FR scheme is substantially different from the FR of Huynh \cite{hu07}. For comparison and completeness, we describe Huynh's flux reconstruction formulation in 2.1, and present the  MMC-FR formulation in 2.2 as follows. 

\subsection{Huynh's flux reconstruction}
Consider a hyperbolic conservation law, Huynh's flux reconstruction consists of the following steps. 
\begin{enumerate}
\item Given the state variable and the flux function at the solution points, compute the primary reconstruction for the flux functions through \eqref{lagintp_f} for all cell  $\Omega_i$, $i=1,2,\cdots,I$.  
\item Compute the flux functions  on two sides of the cell boundary  $x_{i-{1\over2}}$ separately by $f^L_{i-{1\over2}}=f_{i-1}(x_{i-{1\over2}})$ and $f^R_{i-{1\over2}}=f_{i}(x_{i-{1\over2}})$. 
\item Find the numerical flux function, $f^{{\mathcal B}}_{i-{1\over2}}$ at the cell boundary by solving 
\begin{equation}
\begin{split}
 & f^{{\mathcal B}}_{i-{1\over2}}={\rm Riemann} \left( f^L_{i-{1\over2}}, f^R_{i-{1\over2}} \right),
\end{split}
\end{equation}
where ``${\rm Riemann}(\cdot \ ,\ \cdot)$"  denotes a solver for the Riemann problem given the values at the left and right sides of cell boundary  $x_{i-{1\over2}}$.
\item  The modified flux function over $\Omega_i$ in Huynh's formulation is then computed by     
\begin{equation}
\tilde{f}_i(x)= {f}_i(x) + [f^{{\mathcal B}}_{i-{1\over2}}-f_i(x_{i-{1\over2}})]g_{iL}(x) +[f^{{\mathcal B}}_{i+{1\over2}}-f_i(x_{i+{1\over2}})]g_{iR}(x),  
\label{mod_f}
\end{equation} 
where  $g_{iL}(x)$ and  $g_{iR}(x)$ are the so called correction functions for cell $\Omega_i$ which enforce the continuity at the cell boundaries, and satisfies 
\begin{equation}
g_{iL}(x_{i-{1\over2}})=1; \quad g_{iL}(x_{i+{1\over2}})=0   
\label{corre_l}
\end{equation} 
and 
\begin{equation}
g_{iR}(x_{i-{1\over2}})=0; \quad g_{iR}(x_{i+{1\over2}})=1.    
\label{corre_l}
\end{equation} 

Both  $g_{iL}(x)$ and  $g_{iR}(x)$ are polynomials of degree $K$ approximating the zero function, so the modified flux function $\tilde{f}_i(x)$ is of $K$ degree. Shown in \cite{hu07}, the  nodal discontinuous Galerkin (DG) methods \cite{hw08} and spectral collocation method \cite{ko96} can be respectively recovered by choosing the correction function to be the Radau polynomial and the Lagrange polynomial for the Chebyshev collocation points. See \cite{hu07} for an comprehensive and detailed discussions. 

\item As long as the modified flux function $\tilde{f}_i(x)$ is found, the numerical solutions are updated by the semi-discretized equations \eqref{upu} using algorithms for ordinary differential equations, such as the Runge-Kutta scheme.
\end{enumerate}

\subsection{Multi-moment constrained flux reconstruction}

The MMC-FR approach constructs the modified flux function $\tilde{f}_i(x)$ in a different manner as follows.  
\begin{enumerate}
\item Compute the primary reconstruction for the flux functions through \eqref{lagintp_f} for all cell  $\Omega_i$, $i=1,2,\cdots,I$.  
\item Compute the flux functions  on two sides of the cell boundary $x_{i-{1\over2}}$ by $f^L_{i-{1\over2}}=f_{i-1}(x_{i-{1\over2}})$ and $f^R_{i-{1\over2}}=f_{i}(x_{i-{1\over2}})$, and their derivatives by $f^{[m]L}_{xi-{1\over2}}=f^{[m]}_{xi-1}(x_{i-{1\over2}})=\frac{d^m}{dx^m}\left ( f_{i-1}(x_{i-{1\over2}}) \right) $ and $f^{[m]R}_{xi-{1\over2}}=f^{[m]}_{xi}(x_{i-{1\over2}})=\frac{d^m}{dx^m}\left ( f_{i}(x_{i-{1\over2}}) \right)$;
\item Find the flux function, $f^{{\mathcal B}}_{i-{1\over2}}$,  and its derivatives, $f^{ [m]  {\mathcal B}}_{xi-{1\over2}}$, at the cell boundary by solving 
\begin{equation}
\begin{split}
 & f^{{\mathcal B}}_{i-{1\over2}}={\rm Riemann} \left( f^L_{i-{1\over2}}, f^R_{i-{1\over2}} \right), \\
 & f^{ [m]  {\mathcal B}}_{xi-{1\over2}}={\rm DRiemann} \left( f^{[m]L}_{xi-{1\over2}}, f^{[m]R}_{xi-{1\over2}} \right),
\end{split}
\end{equation}
where ``${\rm DRiemann}(\cdot \ ,\ \cdot)$" denotes a solver for the derivative Riemann problem (DRP).  
\item The modified flux function  $\tilde{f}_i(x)$ of $K$ degree is then constructed by properly choosing $K+1$ constraints of two kinds, i.e. a) the continuity conditions of flux function as well as its derivatives at the cell boundaries, referred to as {\it boundary constraints}, and b) the constraint conditions at some points inside the mesh cell which can be directly computed from the primary flux function $f_i(x)$, referred to as {\it  interior constraints}.   
\item 
The modified flux function $\tilde{f}_i(x)$ is then obtained from the  $K+1$ constraint conditions.  Thus, the numerical solutions are updated by solving the semi-discretized equations \eqref{upu} through time integration.
\end{enumerate}

It is obvious that the major difference between the FR of Huynh and the MMC-FR lies in steps from (ii) to (iv). The FR of Huynh only requires the continuity of the flux function at the cell boundaries and all the constraining conditions used in determining the modified flux reconstruction are given by the point values, and the interpolation is essentially of Lagrange type, whereas the MMC-FR requires the continuities at the cell boundaries of not only the flux function itself but also its derivatives, which leads to a Hermite type reconstruction of the modified flux function where both the point values and the derivatives are used as the constraints.

In the MMC-FR formulation, we need compute the derivative Riemann problems at each cell boundary. Practically, the approximate Riemann solvers, such as the local Lax-Friedrich (LLF) flux \cite{shos88} and the Roe's flux \cite{ro81} can be used. 
 The high-order derivative Riemann problems by linear and homogeneous derivative Riemann problems are detailed in \cite{to01,ti02} for the hyperbolic systems. As addressed in \cite{to01}, since the first-instant plays a leading role in the interaction of the two states, the derivative Riemann problems with these simplifications provide a reasonable accuracy. For the Euler equations in fact, our numerical experiments show that the following linearization to the flux functions gives adequate 
accuracy in terms of both numerical error and convergence rate,
\begin{equation}
 f=Au, \ f_x=Au_x, \cdots, \ f^{[m]}_{x}=Au^{[m]}_{x}, \label{flux-linear}
\end{equation}
 where $A$ is the Jacobian matrix obtained by $A=\partial f/\partial u$.
Provided the derivatives of the state variable $u$ from the cell-wise reconstructions for both sides of a discontinuity, one can find the derivative flux of any order at the expense of the conventional Riemann problem.  

In the present study, the reconstruction is carried out in terms of flux function $f$ itself, instead of the conservative variable $u$. For example, the numerical approximations for flux function $f$ and its derivatives at cell boundary $x_{ip}$ can be calculated by
\begin{equation}
\begin{split}
 {f}^{[m]  {\mathcal B}}_{x ip}
 & =\frac{1}{2} \left( f^{[m]-}_{x ip}+f^{[m]+}_{x ip}- R_{ip} |\Lambda_{ip}| R^{-1}_{ip} (u^{[m]+}_{x ip}-u^{[m]-}_{x ip}) \right) \\
 & =\frac{1}{2} \left( f^{[m]-}_{x ip}+f^{[m]+}_{x ip}- R_{ip} {\rm sgn}(\Lambda_{ip}) R^{-1}_{ip} (f^{[m]+}_{x ip}-f^{[m]-}_{x ip}) \right),
\end{split}
\label{flux-recon}
\end{equation}
where we make use of the relation: $u=A^{-1}f=(R \Lambda R^{-1})^{-1}f=R \Lambda^{-1} R^{-1} f$ with $\Lambda$ being the diagonalized matrix of the eigen values.
The eigen matrices $R$ and $R^{-1}$, as well as the eigen values in $\Lambda $ are directly evaluated by the point values at $x_{ip}$.  

\subsection{Implementation of the MMC-FR formulation}
As discussed above, the MMC-FR uses not only the point value of flux function but also its spatial derivatives as the constraints, and thus is more flexible with greater freedom to experiment with.  We in this section presents the general procedure to construct the modified flux with some concrete examples. 

Given $K$ solution points, the primary reconstruction ${f}_i(x)$ is of $K-1$ degrees. The modified flux reconstruction $\tilde {f}_i(x)$ should be at least of $K$ degree to get the solutions of \eqref{upu} $u_{ik}$ of degree  $K-1$. So, we need $K+1$ or more constraints for the reconstruction.  Suppose we use the continuity constraint conditions up to the $k_L$th derivative of the flux function at the left boundary and the  $k_R$th derivative of the flux function at the right boundary, we have the boundary constraints as
\begin{equation}
\begin{split}
 & \tilde {f}^{[m]}_{xi}(x_{i-{1\over2}})=f^{ [m]  {\mathcal B}}_{xi-{1\over2}}, \quad m=0,1,\cdots, k_L; \\
 & \tilde {f}^{[m]}_{xi}(x_{i+{1\over2}})=f^{ [m]  {\mathcal B}}_{xi+{1\over2}}, \quad m=0,1,\cdots, k_R;  
\end{split} \label{b-c}
\end{equation} 
where the flux function is included as the case of $0$th order derivative, i.e. $f(x)=f^{ [0]}_{x}(x)$. 

We leave the rest constraints,  $k_I=K-k_L-k_R+1$ in number,  to be the interior constraints and determined by coinciding the modified flux function with the primary flux function $f_i(x)$ at some points (referred to as {\it  constraint points}) in terms of point value or spatial derivatives, i.e.  
\begin{equation}
\tilde {f}^{[m']}_{xi}(x_{ik'})={f}^{[m']}_{xi}(x_{ik'}),  \quad k'=1,\cdots, k_I;  \quad m'=0,1,\cdots.
\label{i-c}
\end{equation}
Note that the constraint points $x_{k'}$ are not necessarily the same as the solution points $x_{ik}$. 

From the constraint conditions given in \eqref{b-c} and \eqref{i-c}, the $K$ degree modified flux function $\tilde {f}_{i}(x)$, is uniquely determined.

Next, we give some examples of deriving the flux functions by the MMC-FR approach.  For brevity, we use a local coordinate system $\xi \in [-1,1]$ that transforms the real mesh cell $x \in [x_{i-1/2},x_{i+1/2}]$ by 
\begin{equation}
 \xi = 2\frac{x-x_{i-\frac{1}{2}}}{\Delta x_i}-1,
\end{equation}
where $\Delta x_i=x_{i+1/2}-x_{i-1/2}$.

The constraints \eqref{b-c} and \eqref{i-c} are recast correspondingly as 
\begin{equation}
\begin{split}
 & \tilde {f}^{[m]}_{\xi i}(-1)=f^{[m] {\mathcal B}}_{\xi i}(-1)=\left (\frac{d \xi}{dx} \right )_i^{-m} f^{ [m]  {\mathcal B}}_{xi-{1\over2}}, \quad m=0,1,\cdots, k_L; \\
 & \tilde {f}^{[m]}_{\xi i}(1)=f^{[m] {\mathcal B}}_{\xi i}(1)=\left (\frac{d \xi}{dx} \right )_i^{-m} f^{ [m]  {\mathcal B}}_{xi+{1\over2}}, \quad m=0,1,\cdots, k_R;  
\end{split} \label{b-c-local}
\end{equation} 
and 
\begin{equation}
\tilde {f}^{[m']}_{\xi i}(\xi_{ik'})={f}^{[m']}_{\xi i}(\xi_{ik'}),  \quad k'=1,\cdots, k_I;  \quad m'=0,1,\cdots,
\label{i-c-local}
\end{equation}
where $\xi_{ik'}=\xi(x_{ik'})$. 

The time evolution equations \eqref{upu} for updating the solutions are correspondingly    
 \begin{equation}
{{d u_{ik}} \over {d t}}=-\left (\frac{d \xi}{dx} \right )_i\left ( {{d \tilde{f}_i(\xi)} \over {d \xi}}\right )_{ik}, \quad k=1,2,\cdots,K.  
\label{upu-local}
\end{equation}
 
\begin{itemize}

\item {Three-point scheme}

We consider a scheme having three solution points, $\xi_{1}$, $\xi_{2}$ and
$\xi_{3}$, where the solution $u_{ik}$, $k=1,2,3$, are updated every time step. We assume that the flux $f(u)$ is a function of solution $u$. The values of the flux at the corresponding points, $f_{ik}$, $k=1,2,3$, are computed directly. The primary flux function is then built by  
\begin{equation}
f_i(\xi)=\sum^3_{k=1}f_{ik}\phi_{k}(\xi).  
\label{lagintp_f3}
\end{equation}
where 
\begin{equation}
\phi_{k}=\prod_{l=1,l\neq k}^3 \frac{\xi-\xi_k}{\xi_l-\xi_k}
\end{equation} 
is the basis function of the Lagrange interpolation. 

The continuous flux function and its first order derivative at the cell boundaries in \eqref{b-c-local} are computed from steps (i)-(iii) of 2.2. 

The modified flux function is constructed by using boundary constraints up to $k_L=k_R=1$, while no interior constraint is used, $k_I=0$.  That is, the modified flux function of degree 3 is determined by the following constraint conditions, 

\begin{equation}
\left\{
\begin{split}
& \tilde{f}_i(-1)=f^{{\mathcal B}}_i(-1); \\
& \tilde{f}_i(1)=f^{{\mathcal B}}_i(1); \\
& \tilde{f}^{[1]}_{\xi i}(-1)=f^{[1] {\mathcal B}}_{\xi i}(-1); \\
& \tilde{f}^{[1]}_{\xi i}(1)=f^{[1] {\mathcal B}}_{\xi i}(1). 
\end{split} 
\right.
\label{mcv3_1}
\end{equation} 
\eqref{mcv3_1} is a Hermite interpolation to determine the modified flux function which is written in a polynomial form as,
\begin{equation}
\left\{
\begin{split}
\tilde{f}_i(\xi)=& \frac{1}{4}\left (f^{{\mathcal B}}_i(-1)-f^{{\mathcal B}}_i(1) +f^{{\mathcal B}}_{\xi i}(-1)+f^{{\mathcal B}}_{\xi i}(1) \right )\xi^3 \\
                 +& \frac{1}{4}\left ( f^{{\mathcal B}}_{\xi i}(1)-f^{{\mathcal B}}_{\xi i}(-1) \right )\xi^2 \\
+ & \frac{1}{4}\left (3f^{{\mathcal B}}_i(1)-3f^{{\mathcal B}}_i(-1) -f^{{\mathcal B}}_{\xi i}(-1)-f^{{\mathcal B}}_{\xi i}(1) \right )\xi \\
+ & \frac{1}{4}\left (2f^{{\mathcal B}}_i(1)+2f^{{\mathcal B}}_i(-1) +f^{{\mathcal B}}_{\xi i}(-1)-f^{{\mathcal B}}_{\xi i}(1) \right ). 
\end{split} 
\right.
\label{mcv3_intp1}
\end{equation}    

The first order derivative (gradient) of \eqref{mcv3_1} reads then, 
\begin{equation}
\left\{
\begin{split}
\tilde{f}_{\xi i}(\xi)=& \frac{3}{4}\left (f^{{\mathcal B}}_i(-1)-f^{{\mathcal B}}_i(1) +f^{{\mathcal B}}_{\xi i}(-1)+f^{{\mathcal B}}_{\xi i}(1) \right )\xi^2 \\
                 +& \frac{1}{2}\left ( f^{{\mathcal B}}_{\xi i}(1)-f^{{\mathcal B}}_{\xi i}(-1) \right )\xi \\
+ & \frac{1}{4}\left (3f^{{\mathcal B}}_i(1)-3f^{{\mathcal B}}_i(-1) -f^{{\mathcal B}}_{\xi i}(-1)-f^{{\mathcal B}}_{\xi i}(1) \right ). 
\end{split} 
\right.
\label{mcv3_intp1-grd}
\end{equation}    

The derivatives of the modified flux function at the solution points are obtained as  
\begin{equation}
\left\{
\begin{split}
& \left ( {{d \tilde{f}_i(\xi)} \over {d \xi}}\right )_{i1}=\tilde{f}_{\xi i1}=\tilde{f}_{\xi i}(\xi_1); \\
& \left ( {{d \tilde{f}_i(\xi)} \over {d \xi}}\right )_{i2}=\tilde{f}_{\xi i2}=\tilde{f}_{\xi i}(\xi_2); \\
& \left ( {{d \tilde{f}_i(\xi)} \over {d \xi}}\right )_{i3}=\tilde{f}_{\xi i3}=\tilde{f}_{\xi i}(\xi_3).  
\end{split} 
\right.
\label{dflux-3pt}
\end{equation} 
The solutions are then immediately computed by \eqref{upu-local} with a proper time integration algorithm. 

Choosing different solution points results in different schemes. We give two examples next. 

\begin{itemize}
\item {Equidistant points}

If the collocation points are equally spaced and including the cell ends, i.e. $\xi_{1}=-1$, $\xi_{2}=0$ and
$\xi_{3}=1$, we retrieve the third-order MCV scheme\cite{ii09}, where the left/right-most solution points coincide with the cell boundaries. In this case, the continuity conditions of flux function at the cell boundaries are automatically satisfied, and only the derivatives of the flux function need to be computed from the DRP. 

The derivatives of the modified flux function at the solution points are obtained as  
\begin{equation}
\left\{
\begin{split}
& \tilde{f}_{\xi i1}=\tilde{f}_{\xi i}(\xi_1)=f^{{\mathcal B}}_{\xi i}(-1) ; \\
& \tilde{f}_{\xi i2}= \frac{1}{4}\left (3f^{{\mathcal B}}_i(1)-3f^{{\mathcal B}}_i(-1) -f^{{\mathcal B}}_{\xi i}(-1)-f^{{\mathcal B}}_{\xi i}(1) \right ); \\
& \tilde{f}_{\xi i3}=\tilde{f}_{\xi i} (\xi_3)=f^{{\mathcal B}}_{\xi i}(1).  
\end{split} 
\right.
\label{dflux-mcv3_1}
\end{equation} 
It is noted that in the three-point case equidistant points are identical to the Chebyshev-Gauss-Lobatto points. It is straightforward to show the following conservation property,  
\begin{equation}
\int^1_{-1} \tilde{f}_{\xi i}(\xi)d\xi = \sum^3_{k=1}\left( w_{k}\tilde{f}_{\xi ik} \right)=  \frac{1}{3}\tilde{f}_{\xi i1}+  \frac{4}{3}\tilde{f}_{\xi i2}+ \frac{1}{3}\tilde{f}_{\xi i3} = f^{{\mathcal B}}_{i}(1)-f^{{\mathcal B}}_{i}(-1), 
\label{conservation-mcv3-ep}
\end{equation}
where $w_{k}=\int^1_{-1} \phi_k(\xi)d\xi$ are the weights of numerical quadrature. 

\item {Chebyshev-Gauss points}

We use the Chebyshev-Gauss points, i.e. $\xi_{1}=-\sqrt{3}/2$, $\xi_{2}=0$ and
$\xi_{3}=\sqrt{3}/2$, as the solution points. In this case, the continuity conditions of flux function and its derivatives are not necessarily satisfied. So, all of them have to be computed from the DRP. 

The derivatives of the modified flux function at the solution points are obtained as  
\begin{equation}
\left\{
\begin{split}
& \tilde{f}_{\xi i1}= \frac{1}{16}\left (3f^{{\mathcal B}}_i(1)-3f^{{\mathcal B}}_i(-1) +(5+4\sqrt{3})f^{{\mathcal B}}_{\xi i}(-1)+(5-4\sqrt{3})f^{{\mathcal B}}_{\xi i}(1) \right ); \\
& \tilde{f}_{\xi i2}= \frac{1}{4}\left (3f^{{\mathcal B}}_i(1)-3f^{{\mathcal B}}_i(-1) -f^{{\mathcal B}}_{\xi i}(-1)-f^{{\mathcal B}}_{\xi i}(1) \right ); \\
& \tilde{f}_{\xi i3}= \frac{1}{16}\left (3f^{{\mathcal B}}_i(1)-3f^{{\mathcal B}}_i(-1) +(5-4\sqrt{3})f^{{\mathcal B}}_{\xi i}(-1)+(5+4\sqrt{3})f^{{\mathcal B}}_{\xi i}(1) \right ).  
\end{split} 
\right.
\label{dflux-3gauss}
\end{equation} 
Again the numerical conservation yields from  
\begin{equation}
\int^1_{-1} \tilde{f}_{\xi i}(\xi)d\xi = \sum^3_{k=1}\left( w_{k}\tilde{f}_{\xi ik} \right)=  \frac{4}{9}\tilde{f}_{\xi i1}+  \frac{10}{9}\tilde{f}_{\xi i2}+ \frac{4}{9}\tilde{f}_{\xi i3} = f^{{\mathcal B}}_{i}(1)-f^{{\mathcal B}}_{i}(-1).
\label{conservation-mcv3-gp}
\end{equation}

\end{itemize}

\item {Four-point scheme}

When we use four solution points, $\xi_{1}$, $\xi_{2}$,
$\xi_{3}$ and $\xi_{4}$, the primary flux function is constructed by   the Lagrange interpolation, 
\begin{equation}
f_i(\xi)=\sum^4_{k=1}f_{ik}\phi_{k}(\xi).  
\label{lagintp_f4}
\end{equation}

After continuous flux function and its first order derivative at the cell boundaries in \eqref{b-c-local} are computed,  
the modified flux function is constructed by using boundary constraints up to $k_L=k_R=1$, while one interior constraint is used, $k_I=1$.  The modified flux function of degree 4 is determined by the followings, 
\begin{equation}
\left\{
\begin{split}
& \tilde{f}_i(-1)=f^{{\mathcal B}}_i(-1); \\
& \tilde{f}_i(1)=f^{{\mathcal B}}_i(1); \\
& \tilde{f}^{[1]}_{\xi i}(-1)=f^{[1] {\mathcal B}}_{\xi i}(-1); \\
& \tilde{f}^{[1]}_{\xi i}(1)=f^{[1] {\mathcal B}}_{\xi i}(1);\\
& \tilde{f}_{i}(0)=f_i(0). 
\end{split} 
\right.
\label{mcv4_1}
\end{equation} 
\eqref{mcv4_1} is a hybrid expression including both Hermite (the first four) and Lagrange (the last one) interpolations, and can be written in a polynomial form as,
\begin{equation}
\left\{
\begin{split}
\tilde{f}_i(\xi)=& \frac{1}{4}\left (-2f^{{\mathcal B}}_i(-1)-2f^{{\mathcal B}}_i(1) + 4f_{i}(0) -f^{{\mathcal B}}_{\xi i}(-1)+f^{{\mathcal B}}_{\xi i}(1)  \right )\xi^4 \\
                 +& \frac{1}{4}\left ( f^{{\mathcal B}}_i(-1)-f^{{\mathcal B}}_i(1) + f^{{\mathcal B}}_{\xi i}(-1)+f^{{\mathcal B}}_{\xi i}(1)  \right )\xi^3 \\
+ & \frac{1}{4}\left (4f^{{\mathcal B}}_i(-1)+4f^{{\mathcal B}}_i(1) - 8f_{i}(0) +f^{{\mathcal B}}_{\xi i}(-1)-f^{{\mathcal B}}_{\xi i}(1)  \right )\xi^2  \\
+ & \frac{1}{4}\left ( 3f^{{\mathcal B}}_i(1)-3f^{{\mathcal B}}_i(-1) -f^{{\mathcal B}}_{\xi i}(-1)-f^{{\mathcal B}}_{\xi i}(1)  \right )\xi + f_{i}(0).
\end{split} 
\right.
\label{mcv4_intp1}
\end{equation}    

The first order derivative (gradient) of \eqref{mcv4_intp1} reads then, 
\begin{equation}
\left\{
\begin{split}
\tilde{f}_{\xi i}(\xi)=& \left (-2f^{{\mathcal B}}_i(-1)-2f^{{\mathcal B}}_i(1) + 4f_{i}(0) -f^{{\mathcal B}}_{\xi i}(-1)+f^{{\mathcal B}}_{\xi i}(1)  \right )\xi^3 \\
                 +& \frac{3}{4}\left ( f^{{\mathcal B}}_i(-1)-f^{{\mathcal B}}_i(1) + f^{{\mathcal B}}_{\xi i}(-1)+f^{{\mathcal B}}_{\xi i}(1)  \right )\xi^2 \\
+ & \frac{1}{2}\left (4f^{{\mathcal B}}_i(-1)+4f^{{\mathcal B}}_i(1) - 8f_{i}(0) +f^{{\mathcal B}}_{\xi i}(-1)-f^{{\mathcal B}}_{\xi i}(1)  \right )\xi  \\
+ & \frac{1}{4}\left ( 3f^{{\mathcal B}}_i(1)-3f^{{\mathcal B}}_i(-1) -f^{{\mathcal B}}_{\xi i}(-1)-f^{{\mathcal B}}_{\xi i}(1)  \right ). 
\end{split} 
\right.
\label{mcv4_intp1-grd}
\end{equation}

The derivatives of the modified flux function at the solution points are obtained as  
\begin{equation}
\left\{
\begin{split}
& \left ( {{d \tilde{f}_i(\xi)} \over {d \xi}}\right )_{i1}=\tilde{f}_{\xi i1}=\tilde{f}_{\xi i}(\xi_1); \\
& \left ( {{d \tilde{f}_i(\xi)} \over {d \xi}}\right )_{i2}=\tilde{f}_{\xi i2}=\tilde{f}_{\xi i}(\xi_2); \\
& \left ( {{d \tilde{f}_i(\xi)} \over {d \xi}}\right )_{i3}=\tilde{f}_{\xi i3}=\tilde{f}_{\xi i}(\xi_3); \\
& \left ( {{d \tilde{f}_i(\xi)} \over {d \xi}}\right )_{i4}=\tilde{f}_{\xi i4}=\tilde{f}_{\xi i}(\xi_4).  
\end{split} 
\right.
\label{dflux-4pt}
\end{equation} 
The solutions are then immediately computed by \eqref{upu-local} with a proper time integration algorithm. 

An alternative to the constraint conditions \eqref{mcv4_1} is 
\begin{equation}
\left\{
\begin{split}
& \tilde{f}_i(-1)=f^{{\mathcal B}}_i(-1); \\
& \tilde{f}_i(1)=f^{{\mathcal B}}_i(1); \\
& \tilde{f}^{[1]}_{\xi i}(-1)=f^{[1] {\mathcal B}}_{\xi i}(-1); \\
& \tilde{f}^{[1]}_{\xi i}(1)=f^{[1] {\mathcal B}}_{\xi i}(1);\\
& \frac{d^2\tilde{f}_{i}(0)}{d \xi^2}=\frac{d^2{f}_{i}(0)}{d \xi^2}=f^{[2]}_{\xi i}(0). 
\end{split} 
\right.
\label{mcv4_c-2nd}
\end{equation} 
Here, we retain the curvature of the primary interpolation rather than the value at the cell center.  The modified reconstruction for the flux function is then,
\begin{equation}
\left\{
\begin{split}
\tilde{f}_i(\xi)=& \frac{1}{8}\left (f^{[1]{\mathcal B}}_{\xi i}(1)-f^{[1]{\mathcal B}}_{\xi i}(-1) - 2f^{[2]}_{\xi i}(0)  \right )\xi^4 \\
   +& \frac{1}{4}\left ( f^{{\mathcal B}}_i(1)-f^{{\mathcal B}}_i(-1) + f^{[1]{\mathcal B}}_{\xi i}(-1)+f^{[1]{\mathcal B}}_{\xi i}(1)  \right )\xi^3+\frac{1}{2}f^{[2]}_{\xi i}(0)\xi^2\\
+ & \frac{1}{4}\left ( 3f^{{\mathcal B}}_i(1)-3f^{{\mathcal B}}_i(-1) -f^{{\mathcal B}}_{\xi i}(-1)-f^{{\mathcal B}}_{\xi i}(1)  \right )\xi \\
+ & \frac{1}{8}\left ( 4f^{{\mathcal B}}_i(1)+4f^{{\mathcal B}}_i(-1) +f^{{\mathcal B}}_{\xi i}(-1)-f^{{\mathcal B}}_{\xi i}(-1)  -f^{[2]}_{\xi i}(0) \right ).
\end{split} 
\right.
\label{mcv4_c-2nd-intp1}
\end{equation}

The first order derivative (gradient) of \eqref{mcv4_c-2nd-intp1} at the solution points can be directly computed in the same way. We call this scheme the MCV4 with central 2nd derivative constraint, MCV4\_C2D in short.

The formulations discussed above work for different collocation points. We give some examples as follows. 

\begin{itemize}
\item {Equidistant point }

We use 4 point equally spaced points, $\xi_{1}=-1$, $\xi_{2}=-1/3$,  $\xi_{3}=1/3$ and
$\xi_{4}=1$, which results in the 4-order MCV scheme\cite{ii09}, where the left/right-most solution points coincide with the cell boundaries. In this case, the continuity conditions of flux function at the cell boundaries are automatically satisfied, and only the derivatives of the flux function need to be computed from the DRP. 

The derivatives of the modified flux function of the MCV4 scheme \eqref{mcv4_intp1-grd} at the solution points are obtained as  
\begin{equation}
\left\{
\begin{split}
& \tilde{f}_{\xi i1}=f^{{\mathcal B}}_{\xi i}(-1) ; \\
& \tilde{f}_{\xi i2}= \frac{1}{27}\left (2f^{{\mathcal B}}_i(1)-34f^{{\mathcal B}}_i(-1) -8f^{{\mathcal B}}_{\xi i}(-1)-f^{{\mathcal B}}_{\xi i}(1) + 32f^{{\mathcal B}}_i(0)\right ); \\
& \tilde{f}_{\xi i3}= \frac{1}{27}\left (34f^{{\mathcal B}}_i(1)-2f^{{\mathcal B}}_i(-1) -8f^{{\mathcal B}}_{\xi i}(1)-f^{{\mathcal B}}_{\xi i}(-1) - 32f^{{\mathcal B}}_i(0)\right ); \\
& \tilde{f}_{\xi i4}=f^{{\mathcal B}}_{\xi i}(1).  
\end{split} 
\right.
\label{dflux-mcv4_1_edp}
\end{equation} 

For the MCV4\_C2D scheme, the modified flux function \eqref{mcv4_c-2nd-intp1} results in different formula, 
\begin{equation}
\left\{
\begin{split}
& \tilde{f}_{\xi i1}=f^{{\mathcal B}}_{\xi i}(-1) ; \\
& \tilde{f}_{\xi i2}= \frac{1}{27}\left (18f^{{\mathcal B}}_i(1)-18f^{{\mathcal B}}_i(-1) -4f^{{\mathcal B}}_{\xi i}(-1)-5f^{{\mathcal B}}_{\xi i}(1) -8f^{[2]{\mathcal B}}_{\xi i}(0)\right ); \\
& \tilde{f}_{\xi i3}= \frac{1}{27}\left (18f^{{\mathcal B}}_i(1)-18f^{{\mathcal B}}_i(-1) -4f^{{\mathcal B}}_{\xi i}(1)-5f^{{\mathcal B}}_{\xi i}(-1) + 8f^{[2]{\mathcal B}}_{\xi i}(0)\right ); \\
& \tilde{f}_{\xi i4}=f^{{\mathcal B}}_{\xi i}(1).  
\end{split} 
\right.
\label{dflux-mcv4_1-cent-2nd}
\end{equation} 

We can directly prove the following numerical conservation for both \eqref{dflux-mcv4_1_edp} and \eqref{dflux-mcv4_1-cent-2nd}, 
\begin{equation}
\sum^4_{k=1}\left( \tilde{f}_{\xi ik}\int^1_{-1} \phi_{k}(\xi)d\xi \right)=  \frac{1}{4}\tilde{f}_{\xi i1}+  \frac{3}{4}\tilde{f}_{\xi i2}+ \frac{3}{4}\tilde{f}_{\xi i3} + \frac{1}{4}\tilde{f}_{\xi i4}= f^{{\mathcal B}}_{i}(1)-f^{{\mathcal B}}_{i}(-1).
\label{conservation-mcv4-ep}
\end{equation}

\item {Chebyshev-Gauss-Lobatto points}

For Chebyshev-Gauss-Lobatto points,  $\xi_{1}=-1$, $\xi_{2}=-1/2$,  $\xi_{3}=1/2$ and
$\xi_{4}=1$, the derivatives of the modified flux function \eqref{mcv4_intp1-grd} at the solution points are obtained as  
\begin{equation}
\left\{
\begin{split}
& \tilde{f}_{\xi i1}=f^{{\mathcal B}}_{\xi i}(-1) ; \\
& \tilde{f}_{\xi i2}= \frac{1}{16}\left (-3f^{{\mathcal B}}_i(1)-21f^{{\mathcal B}}_i(-1) -3f^{{\mathcal B}}_{\xi i}(-1)+f^{{\mathcal B}}_{\xi i}(1) + 18f^{{\mathcal B}}_i(0)\right ); \\
& \tilde{f}_{\xi i3}= \frac{1}{16}\left (21f^{{\mathcal B}}_i(1)+3f^{{\mathcal B}}_i(-1) -3f^{{\mathcal B}}_{\xi i}(1)+f^{{\mathcal B}}_{\xi i}(-1) - 18f^{{\mathcal B}}_i(0)\right ); \\
& \tilde{f}_{\xi i4}=f^{{\mathcal B}}_{\xi i}(1). 
\end{split} 
\right.
\label{dflux-4cgl}
\end{equation} 
Those for the  MCV4\_C2D scheme can be obtained from \eqref{mcv4_c-2nd-intp1} directly. Again, the following relation of conservation  holds for MCV4 and MCV4\_C2D, 
\begin{equation}
\sum^4_{k=1}\left( \tilde{f}_{\xi ik}\int^1_{-1} \phi_{k}(\xi)d\xi \right)=  \frac{1}{9}\tilde{f}_{\xi i1}+  \frac{8}{9}\tilde{f}_{\xi i2}+ \frac{8}{9}\tilde{f}_{\xi i3} + \frac{1}{9}\tilde{f}_{\xi i4}= f^{{\mathcal B}}_{i}(1)-f^{{\mathcal B}}_{i}(-1).
\label{conservation-mcv4-glp}
\end{equation}

\item {Chebyshev-Gauss points}

If one chooses Chebyshev-Gauss points, the four solution points locate inside the cell element  $\xi_{1}=\cos(7\pi/8)$, $\xi_{2}=\cos(5\pi/8)$,  $\xi_{3}=\cos(3\pi/8)$ and $\xi_{4}=\cos(\pi/8)$, the derivatives of the modified flux function at the solution points can be obtained immediately by \eqref{mcv4_intp1}. For reader's convenience, we write them in a decimal form as   
\begin{equation}
\left\{
\begin{split}
\tilde{f}_{\xi i1}=&-0.160763093f^{{\mathcal B}}_i(1)-0.380433007f^{{\mathcal B}}_i(-1) +0.0635243020f^{{\mathcal B}}_{\xi i}(1)+0.7168057838f^{{\mathcal B}}_{\xi i}(-1) \\
& + 0.541196100f^{{\mathcal B}}_i(0); \\
\tilde{f}_{\xi i2}=&-0.0131164397f^{{\mathcal B}}_i(1)-1.293446526f^{{\mathcal B}}_i(-1)-0.0048660179f^{{\mathcal B}}_{\xi i}(1)-0.2754640679f^{{\mathcal B}}_{\xi i}(-1) \\
& + 1.306562965f^{{\mathcal B}}_i(0); \\
\tilde{f}_{\xi i3}=&1.293446526f^{{\mathcal B}}_i(1)+0.0131164397f^{{\mathcal B}}_i(-1) -0.2754640679f^{{\mathcal B}}_{\xi i}(1)-0.0048660179f^{{\mathcal B}}_{\xi i}(-1) \\
& -1.306562965f^{{\mathcal B}}_i(0); \\
\tilde{f}_{\xi i4}=&0.380433007f^{{\mathcal B}}_i(1)+0.160763093f^{{\mathcal B}}_i(-1) +0.7168057838f^{{\mathcal B}}_{\xi i}(1)+0.0635243020f^{{\mathcal B}}_{\xi i}(-1) \\
& - 0.541196100f^{{\mathcal B}}_i(0). 
\end{split} 
\right.
\label{dflux-4cg}
\end{equation} 
The numerical conservation is verified by the following equality with round off error,
\begin{equation}
\begin{split}
\sum^4_{k=1}\left( \tilde{f}_{\xi ik}\int^1_{-1} \phi_{k}(\xi)d\xi \right) &=  0.2642977396\tilde{f}_{\xi i1}+  0.7357022609\tilde{f}_{\xi i2}+ 0.7357022607\tilde{f}_{\xi i3} + 0.2642977393\tilde{f}_{\xi i4} \\ 
&= f^{{\mathcal B}}_{i}(1)-f^{{\mathcal B}}_{i}(-1).
\label{conservation-mcv4-gp}
\end{split}
\end{equation}
\end{itemize}

\begin{figure}[h]
\begin{center}
\includegraphics[width=7.5cm]{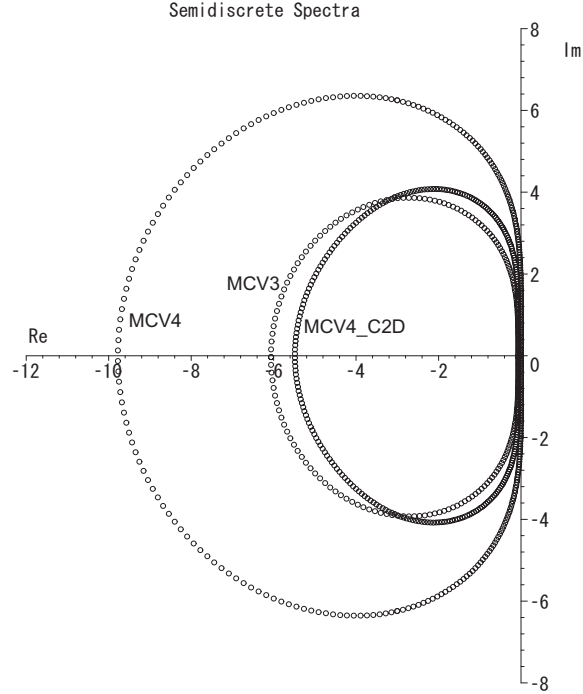} 
\caption{The spectrum of the semi-discrete schemes.  }
\label{sr-3-4} 
\end{center}
\end{figure}

\item {Five-point scheme}

When we use five solution points, $\xi_{1}$, $\xi_{2}$,
$\xi_{3}$, $\xi_{4}$ and $\xi_{5}$, the primary flux function is constructed by the Lagrange interpolation, 
\begin{equation}
f_i(\xi)=\sum^5_{k=1}f_{ik}\phi_{k}(\xi).  
\label{lagintp_f5}
\end{equation}

In this case, we are given more freedom in choosing the constraints to construct the modified flux function. We show two examples as follows. 

\begin{itemize}
\item {MCV5 scheme}

Equidistant solution points are located at $\xi_{k}=-1+2(k-1)/4$, $k=1,2,\cdots, 5$. All constraints are imposed on the cell boundary in terms of the flux derivatives up to $k_L=k_R=2$,
\begin{equation}
\left\{
\begin{split}
& \tilde{f}_i(-1)=f^{{\mathcal B}}_i(-1); \\
& \tilde{f}_i(1)=f^{{\mathcal B}}_i(1); \\
& \tilde{f}^{[1]}_{\xi i}(-1)=f^{[1] {\mathcal B}}_{\xi i}(-1); \\
& \tilde{f}^{[1]}_{\xi i}(1)=f^{[1] {\mathcal B}}_{\xi i}(1);\\
& \tilde{f}^{[2]}_{\xi i}(-1)=f^{[2] {\mathcal B}}_{\xi i}(-1); \\
& \tilde{f}^{[2]}_{\xi i}(1)=f^{[2] {\mathcal B}}_{\xi i}(1). 
\end{split} 
\right.
\label{mcv5_1}
\end{equation} 
\eqref{mcv5_1} is a Hermite interpolation which leads to the next polynomial,
\begin{equation}
\left\{
\begin{split}
\tilde{f}_i(\xi)=& \frac{1}{16}\left (-3f^{{\mathcal B}}_i(-1)+3f^{{\mathcal B}}_i(1) -3f^{[1]{\mathcal B}}_{\xi i}(-1)-3f^{[1]{\mathcal B}}_{\xi i}(1) -f^{[2]{\mathcal B}}_{\xi i}(-1)+f^{[2]{\mathcal B}}_{\xi i}(1)  \right )\xi^5 \\
                 + & \frac{1}{16}\left (f^{[1]{\mathcal B}}_{\xi i}(-1)-f^{[1]{\mathcal B}}_{\xi i}(1) +f^{[2]{\mathcal B}}_{\xi i}(-1)+f^{[2]{\mathcal B}}_{\xi i}(1)  \right )\xi^4  \\
+& \frac{1}{8}\left (5f^{{\mathcal B}}_i(-1)-5f^{{\mathcal B}}_i(1) +5f^{[1]{\mathcal B}}_{\xi i}(-1)+5f^{[1]{\mathcal B}}_{\xi i}(1) +f^{[2]{\mathcal B}}_{\xi i}(-1)-f^{[2]{\mathcal B}}_{\xi i}(1)  \right )\xi^3   \\
+ & \frac{1}{8}\left ( -3f^{[1]{\mathcal B}}_{\xi i}(-1)+3f^{[1]{\mathcal B}}_{\xi i}(1) -f^{[2]{\mathcal B}}_{\xi i}(-1)-f^{[2]{\mathcal B}}_{\xi i}(1)  \right )\xi^2 \\
+& \frac{1}{16}\left (-15f^{{\mathcal B}}_i(-1)+15f^{{\mathcal B}}_i(1) -7f^{[1]{\mathcal B}}_{\xi i}(-1)-7f^{[1]{\mathcal B}}_{\xi i}(1) -f^{[2]{\mathcal B}}_{\xi i}(-1)+f^{[2]{\mathcal B}}_{\xi i}(1)  \right )\xi   \\
+ & \frac{1}{16}\left (8f^{{\mathcal B}}_i(-1)+8f^{{\mathcal B}}_i(1) +5f^{[1]{\mathcal B}}_{\xi i}(-1)-5f^{[1]{\mathcal B}}_{\xi i}(1) +f^{[2]{\mathcal B}}_{\xi i}(-1)+f^{[2]{\mathcal B}}_{\xi i}(1)  \right ). 
\end{split} 
\right.
\label{mcv5_intp1}
\end{equation}    

The derivatives of the modified flux function at the solution points then read  
\begin{equation}
\left\{
\begin{split}
& \tilde{f}_{\xi i1}=f^{[1]{\mathcal B}}_{\xi i}(-1) ; \\
& \tilde{f}_{\xi i2}= \frac{1}{256}\left (135f^{{\mathcal B}}_i(1)-135f^{{\mathcal B}}_i(-1) +81f^{[1]{\mathcal B}}_{\xi i}(-1)-95f^{[1]{\mathcal B}}_{\xi i}(1) +27f^{[2]{\mathcal B}}_{\xi i}(-1)+21f^{[2]{\mathcal B}}_{\xi i}(1) \right ); \\
& \tilde{f}_{\xi i3}= \frac{1}{16}\left (15f^{{\mathcal B}}_i(1)-15f^{{\mathcal B}}_i(-1) -7f^{[1]{\mathcal B}}_{\xi i}(-1)-7f^{[1]{\mathcal B}}_{\xi i}(1) -f^{[2]{\mathcal B}}_{\xi i}(-1)+f^{[2]{\mathcal B}}_{\xi i}(1) \right ); \\
& \tilde{f}_{\xi i4}= \frac{1}{256}\left (135f^{{\mathcal B}}_i(1)-135f^{{\mathcal B}}_i(-1) -95f^{[1]{\mathcal B}}_{\xi i}(-1)+81f^{[1]{\mathcal B}}_{\xi i}(1) -21f^{[2]{\mathcal B}}_{\xi i}(-1)-27f^{[2]{\mathcal B}}_{\xi i}(1) \right ); \\
& \tilde{f}_{\xi i5}=f^{[1]{\mathcal B}}_{\xi i}(1). 
\end{split} 
\right.
\label{dflux-mcv5}
\end{equation} 
The numerical conservation can be obtained from the following relation  
\begin{equation}
\sum^5_{k=1}\left( \tilde{f}_{\xi ik}\int^1_{-1} \phi_{k}(\xi)d\xi \right)=  \frac{1}{45}\left( 7\tilde{f}_{\xi i1}+ 32\tilde{f}_{\xi i2}+ 12\tilde{f}_{\xi i3} + 32\tilde{f}_{\xi i4}+7\tilde{f}_{\xi i5} \right) = f^{{\mathcal B}}_{i}(1)-f^{{\mathcal B}}_{i}(-1).
\label{conservation-mcv5}
\end{equation}

\begin{table}[h]
\begin{center}
\caption{Numerical errors and convergence rate for three-point schemes.}
\begin{tabular}{lcccc} \hline
 Scheme & &$\omega=\pi/8$ & $\omega=\pi/16$ & order \\ \hline 
 MCV3 & &$-3.25\times 10^{-4}-3.33\times 10^{-5}i$  & $-2.06\times 10^{-5}-1.07\times 10^{-6}i$ & 2.99 \vspace{0.1cm} \\ \hline
\end{tabular}\label{error-3ps}                            
\end{center}
\end{table}

\item {MCV5\_PV24 scheme}

We locate the solution at the Chebyshev-Gauss-Lobatto points, $\xi_k=-\cos((k-1)\pi/4$, $k=1,2,\cdots, 5$, and combine the multi-moment constraints at the cell boundary in terms of the flux derivatives up to $k_L=k_R=1$ and two collocation constraints at point $k=2$ and $k=4$ inside the cell,
\begin{equation}
\left\{
\begin{split}
& \tilde{f}_i(-1)=f^{{\mathcal B}}_i(-1); \\
& \tilde{f}_i(1)=f^{{\mathcal B}}_i(1); \\
& \tilde{f}^{[1]}_{\xi i}(-1)=f^{[1] {\mathcal B}}_{\xi i}(-1); \\
& \tilde{f}^{[1]}_{\xi i}(1)=f^{[1] {\mathcal B}}_{\xi i}(1);\\
& \tilde{f}_{i}(\xi_2)=f_i(\xi_2); \\
& \tilde{f}_{i}(\xi_4)=f_i(\xi_4); . 
\end{split} 
\right.
\label{mccc5_1}
\end{equation} 
\eqref{mccc5_1} is a mixture of Hermite interpolation and point collocation (Lagrange interpolation) from which the modified flux function is obtained in a polynomial form as,
\begin{equation}
\left\{
\begin{split}
\tilde{f}_i(\xi)=& \left[ \frac{5}{2}\left (f^{{\mathcal B}}_i(-1)-f^{{\mathcal B}}_i(1)\right)+\frac{1}{2}\left ( f^{[1]{\mathcal B}}_{\xi i}(-1)+f^{[1]{\mathcal B}}_{\xi i}(1)\right)+2\sqrt{2}\left ( -f_i(\xi_2 )+f_i(\xi_4 )  \right ) \right] \xi^5 \\
                 + & \left[ {2}\left (-f^{{\mathcal B}}_i(-1)-f^{{\mathcal B}}_i(1)\right)-\frac{1}{2}\left ( -f^{[1]{\mathcal B}}_{\xi i}(-1)+f^{[1]{\mathcal B}}_{\xi i}(1)\right)+2\left ( f_i(\xi_2 )+f_i(\xi_4 )  \right ) \right] \xi^4  \\
+& \left[ \frac{19}{4}\left (f^{{\mathcal B}}_i(1)-f^{{\mathcal B}}_i(-1)\right)-\frac{3}{4}\left ( f^{[1]{\mathcal B}}_{\xi i}(-1)+f^{[1]{\mathcal B}}_{\xi i}(1)\right)+4\sqrt{2}\left ( f_i(\xi_2 )-f_i(\xi_4 )  \right ) \right] \xi^3   \\
+ & \left[ {4}\left (f^{{\mathcal B}}_i(-1)+f^{{\mathcal B}}_i(1)\right)+\frac{3}{4}\left ( f^{[1]{\mathcal B}}_{\xi i}(-1)-f^{[1]{\mathcal B}}_{\xi i}(1)\right)-4\left ( f_i(\xi_2 )+f_i(\xi_4 )  \right ) \right] \xi^2 \\
+& \left[ \frac{7}{4}\left (f^{{\mathcal B}}_i(-1)-f^{{\mathcal B}}_i(1)\right)+\frac{1}{4}\left ( f^{[1]{\mathcal B}}_{\xi i}(-1)+f^{[1]{\mathcal B}}_{\xi i}(1)\right)+2\sqrt{2}\left ( -f_i(\xi_2 )+f_i(\xi_4 )  \right ) \right] \xi   \\
+ & -\frac{3}{2}\left (f^{{\mathcal B}}_i(-1)+f^{{\mathcal B}}_i(1)\right)+\frac{1}{4}\left ( f^{[1]{\mathcal B}}_{\xi i}(1)-f^{[1]{\mathcal B}}_{\xi i}(-1)\right)+2\sqrt{2}\left ( -f_i(\xi_2 )+f_i(\xi_4 )  \right ) . 
\end{split} 
\right.
\label{mcv5_intp1}
\end{equation}    

The derivatives of the modified flux function at the solution points then read  
\begin{equation}
\left\{
\begin{split}
\tilde{f}_{\xi i1}=&f^{[1]{\mathcal B}}_{\xi i}(-1) ; \\
\tilde{f}_{\xi i2}=&  \left(\frac{9}{4}-2\sqrt{2}\right)f^{{\mathcal B}}_i(1)-\left(\frac{9}{4}+2\sqrt{2}\right)f^{{\mathcal B}}_i(-1)+\frac{1}{4}\left(\sqrt{2}-1\right)f^{[1]{\mathcal B}}_{\xi i}(1)- \frac{1}{4}\left(\sqrt{2}+1\right) f^{[1]{\mathcal B}}_{\xi i}(-1)+ \\
&\frac{\sqrt{2}}{2}\left(7f_i(\xi_2 )+f_i(\xi_4 )\right); \\
\tilde{f}_{\xi i3}= &  \frac{7}{4}\left(f^{{\mathcal B}}_i(-1)-f^{{\mathcal B}}_i(1)\right)+\frac{1}{4}\left(f^{[1]{\mathcal B}}_{\xi i}(1)+ f^{[1]{\mathcal B}}_{\xi i}(-1)\right)+ 2\sqrt{2}\left(f_i(\xi_4 )-f_i(\xi_2 )\right); \\
\tilde{f}_{\xi i4}= &  \left(\frac{9}{4}+2\sqrt{2}\right)f^{{\mathcal B}}_i(1)-\left(\frac{9}{4}-2\sqrt{2}\right)f^{{\mathcal B}}_i(-1)+\frac{1}{4}\left(\sqrt{2}-1\right)f^{[1]{\mathcal B}}_{\xi i}(-1)- \frac{1}{4}\left(\sqrt{2}+1\right) f^{[1]{\mathcal B}}_{\xi i}(1)- \\
&\frac{\sqrt{2}}{2}\left(7f_i(\xi_4 )+f_i(\xi_2 )\right); \\
\tilde{f}_{\xi i5}=&f^{[1]{\mathcal B}}_{\xi i}(1). 
\end{split} 
\right.
\label{dflux-mccc5}
\end{equation} 
The numerical conservation can be obtained from the following relation  
\begin{equation}
\sum^5_{k=1}\left( \tilde{f}_{\xi ik}\int^1_{-1} \phi_{k}(\xi)d\xi \right)=  \frac{1}{15}\left( \tilde{f}_{\xi i1}+ 8\tilde{f}_{\xi i2}+ 4\tilde{f}_{\xi i3} + 8\tilde{f}_{\xi i4}+7\tilde{f}_{\xi i5} \right) = f^{{\mathcal B}}_{i}(1)-f^{{\mathcal B}}_{i}(-1).
\label{conservation-mccc5}
\end{equation}

\item {MCV5\_2D24 scheme}

Another choice is to use the 2nd-order derivatives of the primary interpolation function as the constraints at points $k=2$ and $k=4$, 
\begin{equation}
\left\{
\begin{split}
& \tilde{f}_i(-1)=f^{{\mathcal B}}_i(-1); \\
& \tilde{f}_i(1)=f^{{\mathcal B}}_i(1); \\
& \tilde{f}^{[1]}_{\xi i}(-1)=f^{[1] {\mathcal B}}_{\xi i}(-1); \\
& \tilde{f}^{[1]}_{\xi i}(1)=f^{[1] {\mathcal B}}_{\xi i}(1);\\
& \frac{d^2\tilde{f}_{i}(\xi_2)}{d \xi^2}=\frac{d^2{f}_{i}(\xi_2)}{d \xi^2}=f^{[2]}_{\xi i}(\xi_2);\\
& \frac{d^2\tilde{f}_{i}(\xi_4)}{d \xi^2}=\frac{d^2{f}_{i}(\xi_4)}{d \xi^2}=f^{[2]}_{\xi i}(\xi_4).
\end{split} 
\right.
\label{mccc5_2}
\end{equation} 
Constraint conditions \eqref{mccc5_2} results in another scheme, so called MCV5\_ 2D24. In the same manner, the derivatives at the solution points, $\tilde{f}_{\xi ik}$, $k=1,\cdots,5$, can be directly obtained , and the  numerical conservation can be proven.

\end{itemize}

\end{itemize}

Shown above, the MMC-FR formulation can be implemented in an efficient manner that only involves the nodal values. Higher order schemes can be straightforwardly devised by increasing the solution points and proper constraint conditions. As we will show later, although convergence rate of at least $K$th order can be easily achieved for a $K$-point scheme, the numerical errors and the stable CFL restriction differ by the constraint conditions. Moreover, the numerical conservation can be rigorously guaranteed as long as the modified flux function at cell boundaries are continuous.

\begin{table}[h]
\begin{center}
\caption{Same as Table \ref{error-3ps}, but for four-point schemes. Note that the errors are examined at $\omega=\pi/4$ and $\omega=\pi/8$.}
\begin{tabular}{lcccc} \hline
 Scheme & &$\omega=\pi/4$ & $\omega=\pi/8$ & order  \\ \hline 
 MCV4 & &$-4.91\times 10^{-5}+9.42\times 10^{-5}i$  & $-7.88\times 10^{-7}+3.17\times 10^{-6}i$ & 4.02 \vspace{0.1cm} \\ 
 MCV4\_C2D & &$-1.95\times 10^{-4}+5.71\times 10^{-4}i$  & $-3.15\times 10^{-6}+1.91\times 10^{-5}i$ & 3.96 \vspace{0.1cm} \\ \hline
\end{tabular}  \label{error-4ps}                           
\end{center}
\end{table}

\section{Fourier analysis}

In this section, we evaluated the numerical schemes previously discussed by examining the Fourier mode transported with the following advection equation, 
\begin{equation}
{{\partial u} \over {\partial t}}+{{\partial u} \over {\partial x}} =0.  
\label{1dade}
\end{equation} 
The theoretical tools used hereafter are mainly developed in \cite{hir07, hu07}. 

We use a wave solution,  
\begin{equation}
u(x,t)=e^{I\omega(x+t)}, 
\label{e-sol}
\end{equation}
 and represent it on a grid whose $i$th cell is defined by $[i-1/2,i+1/2]$ where the solution points $x_{ik}$, $k=1,2,\cdots,K$, are located. The solutions are then  $u_{ik}=e^{I\omega(i+\xi_k/2)}$. We consider a scheme constructed over cell $i$ with the information from its upwinding cell $(i-1)$. Recall that  boundary flux and its derivatives, $f^{{\mathcal B}}$ and $f^{[m]{\mathcal B}}_{x }$, are completely upwinding and $f=u$ in this particular case, the schemes discussed above are summarized by 
\begin{equation}
\left(\begin{array}{c}
\displaystyle \frac{d u_{i1}}{dt}\\
\displaystyle \frac{d u_{i2}}{dt}\\
\cdots\\
\displaystyle \frac{d u_{iK}}{dt}
\end{array}\right)
= \left( \begin{array}{cccccccc}
c_{(i-1)11} & c_{(i-1)12} & \cdots & c_{(i-1)1K} & c_{i11} & c_{i12} & \cdots & c_{i1K} \\
c_{(i-1)21} & c_{(i-1)22} & \cdots & c_{(i-1)2K} & c_{i21} & c_{i22} & \cdots & c_{i2K} \\
\cdots  & \cdots  & \cdots & \cdots  & \cdots  & \cdots  & \cdots & \cdots  \\
c_{(i-1)K1} & c_{(i-1)K2} & \cdots & c_{(i-1)KK} & c_{iK1} & c_{iK2} & \cdots & c_{iKK} \\
\end{array} \right)
\left( \begin{array}{c}
u_{(i-1)1}  \\
u_{(i-1)2}  \\
\cdots  \\
u_{(i-1)K}   \\
u_{i1}  \\
u_{i2}  \\
\cdots  \\
u_{iK}  \end{array} \right)
\label{m-form}
\end{equation}

From the wave solution, we have 
$u_{ik}=e^{I\omega(i+\xi_k/2)}$ and  $u_{(i-1)k}=e^{-I\omega} u_{ik}$, which reduces \eqref{m-form} into 
\begin{equation}
\left(\begin{array}{c}
\displaystyle \frac{d u_{i1}}{dt}\\
\displaystyle \frac{d u_{i2}}{dt}\\
\cdots\\
\displaystyle \frac{d u_{iK}}{dt}
\end{array}\right)
= \left( \begin{array}{cccccccc}
c_{(i-1)11}e^{-I\omega}+ c_{i11} & c_{(i-1)12}e^{-I\omega} +c_{i12} & \cdots & c_{(i-1)1K}e^{-I\omega} + c_{i1K}  \\
c_{(i-1)21}e^{-I\omega}+ c_{i21} & c_{(i-1)22}e^{-I\omega} +c_{i22} & \cdots & c_{(i-1)2K}e^{-I\omega} + c_{i2K}  \\
\cdots  & \cdots  & \cdots & \cdots   \\
c_{(i-1)K1}e^{-I\omega}+ c_{iK1} & c_{(i-1)K2}e^{-I\omega} +c_{iK2} & \cdots & c_{(i-1)KK}e^{-I\omega} + c_{iKK}  \\
\end{array} \right)
\left( \begin{array}{c}
u_{i1}  \\
u_{i2}  \\
\cdots  \\
u_{iK}  \end{array} \right), 
\label{m-form-r}
\end{equation}
or 
\begin{equation}
 \frac{d \mathbf{u}_{i}}{dt}=\mathbf{S}\mathbf{u}_{i}. 
\label{m-form-rv}
\end{equation}

As discussed in \cite{hu07}, the constraint conditions make the essential difference in matrix  $\mathbf S$, while different arrangements of the solution points result in similar matrices which have the same eigenvalues. In the following discussions, equi-distanced points are used for the MMC-FR schemes.  

\begin{figure}[h]
\begin{center}
\includegraphics[width=8.5cm]{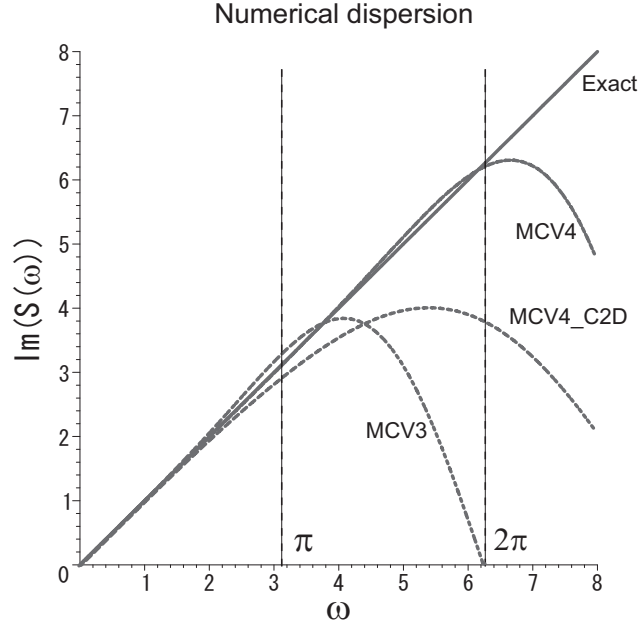} 
\caption{Numerical dispersion relations for the schemes. }
\label{disper-3} 
\end{center}
\end{figure} 

The properties of the numerical schemes can be examined by analyzing the eigenvalues of \eqref{m-form-rv}. Fig.\ref{sr-3-4} shows the spectrum (collection of all eigen values) of $\mathbf S$ for different schemes. It is observed that all eigenvalues lie on the left half of the real axis, i.e the negative real parts indicate that all the schemes are stable under the CFL conditions. The allowable CFL numbers for computational stability can be estimated by the largest eigenvalue, the spectral radius $\rho$ for each scheme, i.e. a scheme has a larger spectral radius has to use a smaller CFL number for computational stability. We know from Fourier analysis that  $\rho_{\rm MCV3}=6.0$, $\rho_{MCV4}=9.78$ and $\rho_{MCV4\_C2D}=5.42$.  The MCV4\_C2D scheme has a spectral radius even smaller than the three-point MCV3 scheme. Shown later, our numerical tests for pure advection equation verify the observations from the spectral radius analysis presented here.  

\begin{figure}[h]
\begin{center}
 \hspace{0.0cm} \includegraphics[width=11.5cm]{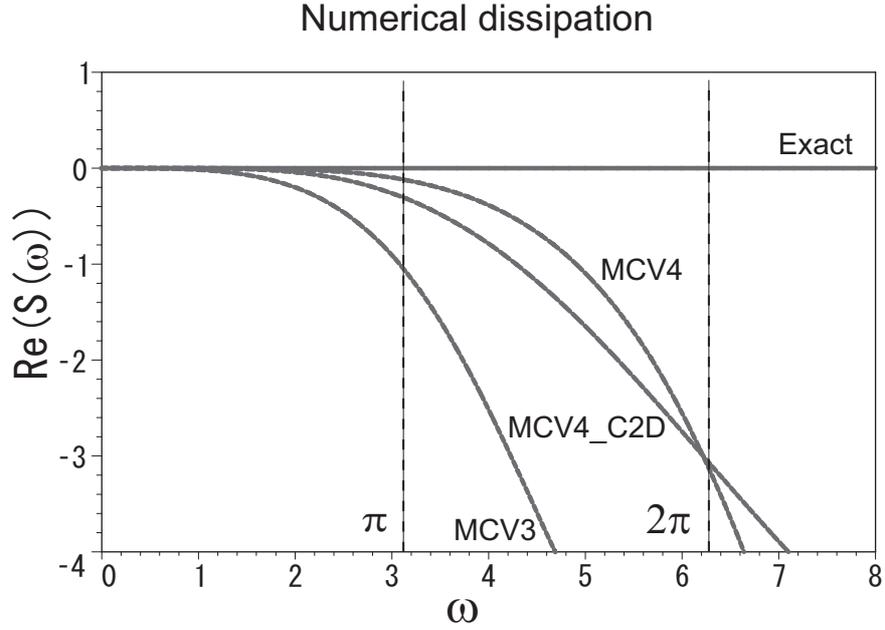} 
\caption{The numerical dissipation relations for the schemes. }
\label{dissip-3} 
\end{center}
\end{figure} 

The numerical errors of different schemes can be examined by comparing the principal eigenvalue of  $\mathbf S$, $\lambda^p_{\mathbf S(\omega)}$, with the exact solution, $-I\omega$, of the advection equation \eqref{1dade} for initial condition,  
\begin{equation}
u(x,0)=e^{I\omega x}. 
\label{ic}
\end{equation}
The error of a given semi-discrete formulation is 
\begin{equation}
E(\omega)=\lambda^p_{\mathbf S(\omega)}+I\omega, 
\label{err-def}
\end{equation}
and the convergence rate is evaluated by 
\begin{equation}
m=\left(\ln\left(\frac{E(\omega)}{E(\omega/2)}\right)/\ln(2)\right )-1. 
\label{err-def}
\end{equation}

The numerical errors of the MCV3 scheme are given in Table \ref{error-3ps}. The MCV3 scheme has a 3rd-order convergence rate. The errors of the four-point schemes are given in Table \ref{error-4ps}. The two MMC-FR schemes, MCV4 and MCV4\_C2D, are stable and of 4th-order convergence rate. 

The  dispersion and dissipation relations of the spatial discretizations can be evaluated by plotting the real and imaginary parts of the principal eigenvalues as a functions of the wave number $\omega$. From Fig.\ref{disper-3}, we find that all schemes agree well with the exact solution for $\omega \le 4$, which reveals a superior numerical dispersion compared to conventional finite difference method or finite volume method. For the four-point schemes, the MCV4 scheme adequately recovers the dispersion relation for all waves of $\omega \le 7$. The MCV4\_C2D scheme, however, is less accurate  for high wave numbers. Fig.\ref{dissip-3} shows that both MCV4 and MCV4\_C2D are more accurate in numerical dissipation even for high wave number.

We evaluated the convergence rates of the schemes discussed above by solving the linear scalar
equation \eqref{1dade} on gradually refined grids with the smooth
initial condition defined by
\begin{equation}\label{eq:test_initial}
    u(x,t=0)=\sin (\pi x). 
\end{equation}
A periodic boundary conditions are specified. The normalized errors
$\ell_1$, $\ell_2$ and $\ell_{\infty}$ at $t=2$ are given in Table \ref{converg-rates}. 

All schemes get the expected convergence rates. The numerical errors, however, varies according to the constraint conditions imposed.  It is observed that the constraints on the point values at the interior points will reduce the numerical errors, whereas the constraints in terms of the second order derivatives result in larger numerical errors. 

\begin{table}[h]
\caption{Normalized errors and convergence rates of the MCV type schemes.} {\small
\begin{center}
\begin{tabular}{lccccccc}
\hline
\hline MCV3 & Mesh & $\ell_1$ & $\ell_1$-order & $\ell_2$ & $\ell_2$-order  & $\ell_{\infty}$ & $\ell_{\infty}$-order \\
\hline  &10        &  $1.29\times10^{-2}$ &   -  & $1.43\times10^{-2}$ &   -  & $2.03\times10^{-2}$ &  -   \\
           &20        &  $1.69\times10^{-3}$ & 2.93 & $1.87\times10^{-3}$ & 2.93 & $2.69\times10^{-3}$ & 2.92 \\
           &40        &  $2.14\times10^{-4}$ & 2.98 & $2.37\times10^{-4}$ & 2.98 & $3.37\times10^{-4}$ & 3.00 \\
           &80        &  $2.68\times10^{-5}$ & 3.00 & $2.98\times10^{-5}$ & 2.99 & $4.22\times10^{-5}$ & 3.00 \\
\hline 
\hline MCV4 & Mesh & $\ell_1$ & $\ell_1$-order & $\ell_2$ & $\ell_2$-order  & $\ell_{\infty}$ & $\ell_{\infty}$-order \\
\hline  &10        &  $2.06\times10^{-4}$ &   -  & $2.26\times10^{-4}$ &   -  & $3.24\times10^{-4}$ &  -   \\
           &20        &  $1.31\times10^{-5}$ & 3.98 & $1.46\times10^{-5}$ & 3.95 & $2.09\times10^{-5}$ & 3.95 \\
           &40        &  $8.32\times10^{-7}$ & 3.98 & $9.25\times10^{-7}$ & 3.98 & $1.31\times10^{-6}$ & 4.00 \\
           &80        &  $5.24\times10^{-8}$ & 3.99 & $5.82\times10^{-8}$ & 3.99 & $8.25\times10^{-8}$ & 3.99 \\
\hline
\hline MCV4\_C2D & Mesh & $\ell_1$ & $\ell_1$-order & $\ell_2$ & $\ell_2$-order  & $\ell_{\infty}$ & $\ell_{\infty}$-order \\
\hline  &10        &  $1.22\times10^{-3}$ &   -  & $1.33\times10^{-3}$ &   -  & $1.91\times10^{-3}$ &  -   \\
           &20        &  $7.84\times10^{-5}$ & 3.96 & $8.75\times10^{-5}$ & 3.93 & $1.26\times10^{-4}$ & 3.92 \\
           &40        &  $5.00\times10^{-6}$ & 3.97 & $5.56\times10^{-6}$ & 3.98 & $7.90\times10^{-6}$ & 4.00 \\
           &80        &  $3.15\times10^{-7}$ & 3.99 & $3.50\times10^{-7}$ & 3.99 & $4.96\times10^{-7}$ & 3.99 \\
\hline
\hline MCV5 & Mesh & $\ell_1$ & $\ell_1$-order & $\ell_2$ & $\ell_2$-order  & $\ell_{\infty}$ & $\ell_{\infty}$-order \\
\hline  &10        &  $5.21\times10^{-5}$ &   -  & $5.72\times10^{-5}$ &   -  & $8.18\times10^{-5}$ &  -   \\
           &20        &  $1.67\times10^{-6}$ & 4.96 & $1.85\times10^{-6}$ & 4.95 & $2.65\times10^{-6}$ & 4.95 \\
           &40        &  $5.28\times10^{-8}$ & 4.98 & $5.86\times10^{-8}$ & 4.98 & $8.31\times10^{-8}$ & 5.00 \\
           &80        &  $1.65\times10^{-9}$ & 5.00 & $1.84\times10^{-9}$ & 4.99 & $2.61\times10^{-9}$ & 4.99 \\
\hline 
\hline MCV5\_2D24 & Mesh & $\ell_1$ & $\ell_1$-order & $\ell_2$ & $\ell_2$-order  & $\ell_{\infty}$ & $\ell_{\infty}$-order \\
\hline  &10        &  $4.58\times10^{-5}$ &   -  & $5.02\times10^{-5}$ &   -  & $7.20\times10^{-5}$ &  -   \\
           &20        &  $1.46\times10^{-6}$ & 4.97 & $1.62\times10^{-6}$ & 4.95 & $2.32\times10^{-6}$ & 4.96 \\
           &40        &  $4.63\times10^{-8}$ & 4.99 & $5.13\times10^{-8}$ & 4.98 & $7.29\times10^{-8}$ & 4.99 \\
           &80        &  $1.46\times10^{-9}$ & 4.99 & $1.62\times10^{-9}$ & 4.98 & $2.29\times10^{-9}$ & 4.99 \\
\hline
\hline MCV5\_PV24 & Mesh & $\ell_1$ & $\ell_1$-order & $\ell_2$ & $\ell_2$-order  & $\ell_{\infty}$ & $\ell_{\infty}$-order \\
\hline  &10        &  $3.48\times10^{-6}$ &   -  & $3.85\times10^{-6}$ &   -  & $5.48\times10^{-6}$ &  -   \\
           &20        &  $1.07\times10^{-7}$ & 5.02 & $1.18\times10^{-7}$ & 5.03 & $1.70\times10^{-7}$ & 5.01 \\
           &40        &  $3.33\times10^{-9}$ & 5.01 & $3.70\times10^{-9}$ & 5.00 & $5.25\times10^{-9}$ & 5.02 \\
           &80        &  $1.06\times10^{-10}$ & 4.97 & $1.17\times10^{-10}$ & 4.98 & $1.66\times10^{-10}$ & 4.98 \\
\hline \label{converg-rates}
\end{tabular}
\end{center}
}
\end{table}

The largest allowable CFL numbers for different schemes are also evaluated through the numerical tests for advection equation with a third order Runge-Kutta time integration scheme. The results are shown in Table \ref{max-cfl}. It is found that using the point values of the primary interpolation at the interior points tends to reduce the maximum stable CFL number, while using the 2nd-order derivatives can increase the stable CFL number.  Consistent with the observation from Fig.\ref{sr-3-4} which show a smaller spectral radius of MCV4\_C2D compared to MCV3, the 4th-order scheme MCV4\_C2D can use larger CFL number than the 3rd-order scheme MCV3.  

\begin{table}[h]
\caption{Largest allowable CFL numbers.} {\small
\begin{center}
\begin{tabular}{lcccccc}
\hline 
\hline 
Schemes& MCV3 & MCV4  & MCV4\_C2D &MCV5  & MCV5\_2D24 & MCV5\_PV24 \\
\hline
CFL(max) & 0.425 & 0.275 & 0.485 &0.21  & 0.235 & 0.165 \\
\hline \label{max-cfl}
\end{tabular}
\end{center}
}
\end{table}

\section{Concluding remarks}
We have presented a general formulation using the multi-moment constraints to construct high-order schemes for 
hyperbolic conservation laws. The formulation accommodates a wide range of schemes under the unified framework of flux reconstruction.  

Leaving the practical schemes to be further explored for specified applications, we give some analysis in this paper to show the basic feature of the multi-moment constrained schemes, and obtain the following observations:
\begin{itemize}
\item The multi-moment constrained finite volume schemes presented in \cite{ii09} are stable and possess numerical accuracy superior to the conventional finite volume method. 
\item The numerical conservation can be rigorously guaranteed as long as the modified flux function at a cell boundary is shared by the two neighboring cells.
\item The location of the solution points is not sensitive to the numerical results.
\item Constraints of the point values at interior points improve numerical accuracy, but tends to suffer a more restrictive CFL condition for computational stability. 
\item Constraints in terms of the 2nd-order derivatives, i.e the curvature of the primary reconstruction, greatly relieve CFL restriction for computational stability. It is possible to construct schemes of higher order convergence rate that allow larger stable CFL number.   

\end{itemize}

\section*{Acknowledgment}
This work was supported in part by JSPS KAKENHI Grant Numberi24560187).

\end{document}